\documentclass[10pt,pra,aps,showpacs,twocolumn]{revtex4}
\usepackage{bbm}

\begin{document}

\title{Adiabatic Quantum Computation with a 1D projector Hamiltonian}         
\author{Avatar Tulsi \\
        {\small Department of Physics, Indian Institute of Science, Bangalore-560012, India}}  
    
\email{tulsi9@gmail.com}

\begin{abstract}
Adiabatic quantum computation is based on the adiabatic evolution of quantum systems. We analyse a particular class of qauntum adiabatic evolutions where either the initial or final Hamiltonian is a one-dimensional projector Hamiltonian on the corresponding ground state. The minimum energy gap which governs the time required for a successful evolution is shown to be proportional to the overlap of the ground states of the initial and final Hamiltonians. We show that such evolutions exhibit a rapid crossover as the ground state changes abruptly near the transition point where the energy gap is minimum. Furthermore, a faster evolution can be obtained by performing a partial adiabatic evolution within a narrow interval around the transition point. These results generalize and quantify earlier works.    
\end{abstract}

\pacs{03.67.Lx}

\maketitle

\section{Introduction}

Quantum adiabatic evolution starts with the ground state $|s\rangle$ of the initial Hamiltonian $\mathsf{H}_{s}$ in an $N$-dimensional Hilbert space, and evolves it slowly enough to the ground state $|t\rangle$ of the final Hamiltonian $\mathsf{H}_{t}$. The evolution uses the time-dependent Hamiltonian
\begin{equation}
\mathsf{H}_{\mu} = (1-\mu)\mathsf{H}_{s} + \mu\mathsf{H}_{t}\ ,\ \ \mu \in [0,1].  \label{evolve}
\end{equation} 
The parameter $\mu$ is a function of the time $\tau$. The quantum adiabatic theorem~\cite{messiah} bounds the total evolution time $\Gamma$ required for a successful evolution. Let the eigenspectrum and the excitation gap of $\mathsf{H}_{\mu}$ be
\begin{eqnarray}
&\mathsf{H}_{\mu}|E_{k,\mu}\rangle = E_{k,\mu}|E_{k,\mu}\rangle,& \nonumber \\ 
&E_{0,\mu} \leq E_{1,\mu} \leq \cdots \leq E_{N-1,\mu}\ ,&  \nonumber \\
&g_{\mu} = E_{1,\mu} - E_{0,\mu}\ .& \label{eigendefine}
\end{eqnarray}
The adiabatic theorem states that one can reach $|t\rangle$ with probability close to $1$, when
\begin{equation}
\Gamma  \geq \Theta\left(g_{\rm min}^{-2}\|\mathsf{H}_{s}-\mathsf{H}_{t}\|\right),\ \ g_{\rm min} = \textstyle \min_{\mu} g_{\mu}\ . \label{adiabatictime}
\end{equation}
Conventionally, the Hamiltonians are normalized such that $\|\mathsf{H}_{s} -\mathsf{H}_{t}\|= \Theta(1)$, and $\Gamma$ is bounded from below essentially by $g_{\rm min}^{-2}$. Thus the knowledge of the minimum energy gap $g_{\rm min}$ is essential to determine the minimum time for successful evolution. In general, estimating $g_{\rm min}$ is not an easy task but it can be estimated for some special cases~\cite{farhiqcae,special1,special2,special3,znidaric,rolandlocal}.

In this paper, we analyze a special case when $\mathsf{H}_{t} = -|t\rangle\langle t|$ is a one-dimensional projector Hamiltonian on its ground state $|t\rangle$. Such kind of projector Hamiltonians naturally appears in solutions to decision problems. In Section II, we analyse the eigenspectrum of $\mathsf{H}_{\mu}$ with $\mathsf{H}_{t} = -|t\rangle\langle t|$. We show, under certain assumptions regarding the eigenspectrum of $\mathsf{H}_{s}$, that $g_{\rm min}$ scales as the overlap $\alpha = |\langle s|t\rangle|$ of the ground states of $\mathsf{H}_{s}$ and $\mathsf{H}_{t}$. We also derive the expression for $\mu_{\rm min}$ where $g_{\mu} = g_{\rm min}$. For $g_{\rm min} = O(\alpha)$, (\ref{adiabatictime}) implies that $\Gamma = \Omega(\alpha^{-2})$. In Section III, we show that the ground state of $\mathsf{H}_{\mu}$ evolves significantly only within a narrow interval $[\mu^{-},\mu^{+}]$ around $\mu_{\rm min}$. Exploiting this property, we present a partial adiabatic evolution algorithm with the time complexity $\Gamma' = \Omega(\alpha^{-1})$, which is faster than the standard adiabatic evolution. In Section IV, we conclude by discussing the relation of our work to earlier works on this subject.

\section{Minimum energy gap}

To calculate the minimum energy gap, we first find the eigenspectrum of $\mathsf{H}_{\mu}$. With $\mathsf{H}_{t} = -|t\rangle\langle t|$ in (\ref{evolve}), we have
\begin{equation}
\mathsf{H}_{\mu} = (1-\mu)\mathsf{H}_{s}-\mu|t\rangle\langle t|.  \label{sevolve}
\end{equation}
The eigenspectrum of above Hamiltonian can be analyzed in a similar way as the eigenspectrum of corresponding unitary operator was analyzed in~\cite{tulsi}. We work in the eigenbasis of $\mathsf{H}_{s}$, chosen such that 
\begin{equation}
\langle \ell|\mathsf{H}_{s}|\ell\rangle = \xi_{\ell}\ ,\ 0 = \xi_{0} \leq \xi_{1} \leq \cdots \leq \xi_{N-1}\ . \label{Hseigendefine}
\end{equation}
For simplicity, we consider $|s\rangle \equiv |\ell = 0\rangle$ to be the non-degenerate ground state of $\mathsf{H}_{s}$. We make the following assumptions regarding the eigenspectrum of $\mathsf{H}_{s}$: 
\begin{equation}
|\langle s|t\rangle| \equiv \alpha \ll \xi_{1}\ ,\ \xi_{1}/\xi_{N-1} \not\ll 1\ ,\ \|\mathsf{H}_{s}\| = \xi_{N-1} \not\gg 1. \label{firstassumption}
\end{equation}
The first one can always be satisfied by appropriately scaling $\mathsf{H}_{s}$ (and hence $\xi_{1}$). The time needed to distinguish the ground state of $\mathsf{H}_{s}$ from the excited states, $\Omega(1/\xi_{1})$, is then much smaller than the time scale of the algorithm, $\Gamma=\Theta(1/\alpha)$. The second one constrains $\|\mathsf{H}_{s}\|$ relative to the initial excitation gap $\xi_{1}$, and the third one constrains $\|\mathsf{H}_{s}\|$. 

Let $|E_{k,\mu}\rangle$ be the normalized eigenvectors of $\mathsf{H}_{\mu}$ with eigenvalues $E_{k,\mu}$. We have
\begin{displaymath}
\mathsf{H}_{\mu}|E_{k,\mu}\rangle = E_{k,\mu}|E_{k,\mu}\rangle = [(1-\mu)\mathsf{H}_{s}-\mu|t\rangle\langle t|]|E_{k,\mu}\rangle. 
\end{displaymath} 
Left multiplication by $\langle \ell|$ and $\langle \ell|\mathsf{H}_{s} = \xi_{\ell}\langle \ell|$ gives
\begin{eqnarray}
E_{k,\mu} \langle \ell |E_{k,\mu}\rangle &=& (1-\mu)\langle \ell |\mathsf{H}_{s}|E_{k,\mu}\rangle - \mu \langle \ell |t\rangle \langle t|E_{k,\mu}\rangle \nonumber \\ 
& = & \xi_{\ell} (1-\mu) \langle \ell |E_{k,\mu}\rangle - \mu \langle \ell |t\rangle \langle t|E_{k,\mu}\rangle.  \nonumber
\end{eqnarray}
Thus
\begin{equation}
\langle \ell |E_{k,\mu}\rangle = \mu \frac{\langle \ell|t\rangle \langle t|E_{k,\mu}\rangle}{\xi_{\ell}(1-\mu)-E_{k,\mu}}\ .   \label{lekuoverlap}
\end{equation}
It gives
\begin{eqnarray}
\langle t|E_{k,\mu}\rangle &=& \sum_{\ell} \langle t |\ell\rangle\langle \ell|E_{k,\mu}\rangle \nonumber \\
&=& \mu \langle t|E_{k,\mu}\rangle\sum_{\ell}\frac{|\langle t|\ell\rangle|^{2}}{\xi_{\ell}(1-\mu)-E_{k,\mu}}\ ,  \label{tekuoverlap}
\end{eqnarray}
and we find the secular equation for $\mathsf{H}_{\mu}$ to be
\begin{equation}
\sum_{\ell}\frac{|\langle \ell|t\rangle|^{2}}{\xi_{\ell}(1-\mu)-E_{k,\mu}} = \frac{1}{\mu}\ . \label{adiabaticsecular}
\end{equation}
Since $\xi_{\ell} \geq 0$, the L.H.S. of above equation decreases monotonically as $E_{k,\mu}$ decreases from $0$ to $-\infty$. On the other hand, the R.H.S. is fixed, so the equation can have at most one negative solution for $E_{k,\mu}$. We will see that above equation has a unique negative solution, which is obviously the ground state energy $E_{0,\mu}$ of\ $\mathsf{H}_{\mu}$. 

We assume that the two lowest solutions of (\ref{adiabaticsecular}) obey 
\begin{equation}
|E_{k,\mu}| \ll (1-\mu)\xi_{1}\ .  \label{assumption}
\end{equation}
To find them, we Taylor expand the $\ell \neq 0$ contribution in (\ref{adiabaticsecular}) and ignore $O(E_{k,\mu}^{2})$ terms. That results in the quadratic equation,
\begin{equation}
\alpha^{2}E_{k,\mu}^{-1} - A_{\mu} - B_{\mu}^{2}E_{k,\mu} =  0\ , \label{adiabaticquadratic} 
\end{equation}
yielding two solutions consistent with $|E_{k,\mu}| \ll (1-\mu)\xi_{1}$. The coefficients $A_{\mu},B_{\mu}$ are 
\begin{equation}
A_{\mu} = \frac{\Upsilon_{1}}{1-\mu}-\frac{1}{\mu}\ ,\ \ B_{\mu} = \frac{\sqrt{\Upsilon_{2}}}{1-\mu}\ , \label{adiabaticcoeff}
\end{equation}
where
\begin{equation}
\Upsilon_{p} = \sum_{\ell \neq 0}\frac{|\langle \ell|t\rangle|^{2}}{\xi_{\ell}^{p}}\ ,\ \ p \in \{1,2\}. \label{adiabaticOmega}
\end{equation}
We note the bounds $\xi_{N-1}^{-p} \leq \Upsilon_{p} \leq \xi_{1}^{-p}$, arising from $\sum_{\ell}|\langle \ell|t\rangle|^{2} = 1$. Also, putting $x_{\ell} = |\langle \ell|t\rangle|$ and $y_{\ell} = |\langle \ell|t\rangle|/\xi_{\ell}$ in the Cauchy-Schwartz inequality $(\sum x_{\ell}y_{\ell})^{2} \leq \sum x_{\ell}^{2}\sum y_{\ell}^{2}$, we get 
\begin{equation}
\Upsilon_{1}^{2} \leq \sum_{\ell \neq 0}|\langle \ell|t\rangle|^{2}\Upsilon_{2} \leq \Upsilon_{2}\ . \label{Cauchy}
\end{equation}

The two solutions $E_{\pm,\mu}$ of (\ref{adiabaticquadratic}) have the product $E_{+,\mu}E_{-,\mu} = -\alpha^{2}/B_{\mu}^{2}$. Hence
\begin{equation}
E_{\pm,\mu} = \pm \frac{\alpha}{B_{\mu}}(\tan \eta_{\mu})^{\pm 1} = \pm\frac{\alpha(1-\mu)}{\sqrt{\Upsilon_{2}}}(\tan \eta_{\mu})^{\pm 1}\ .  \label{adiabaticsolutions}
\end{equation}
The sum of the two roots determines the angle $\eta$. We have $E_{+,\mu}+E_{-,\mu} = -A_{\mu}/B_{\mu}^{2} = -(2\alpha/B_{\mu})\cot 2\eta_{\mu}$. Thus
\begin{equation}
\cot 2\eta_{\mu} = \frac{A_{\mu}}{2\alpha B_{\mu}} = \frac{1}{2\alpha \sqrt{\Upsilon_{2}}}\left(\Upsilon_{1}-\frac{1-\mu}{\mu}\right), \label{adiabaticeta}
\end{equation}
with $\eta_{\mu}\in [0,\frac{\pi}{2}]$. As $\eta_{\mu}$ is positive, $E_{-,\mu}$ is indeed the unique negative solution of (\ref{adiabaticsecular}) and hence the ground state energy $E_{0,\mu}$ of $\mathsf{H}_{\mu}$, while $E_{+,\mu}$ is the first excited state energy $E_{1,\mu}$ of $\mathsf{H}_{\mu}$. 

With $\Upsilon_{1} > 0$, let us define the crossover point $\mu^{*}$, and deviation from it $\varepsilon$ as
\begin{equation}
\frac{1-\mu^{*}}{\mu^{*}} =  \Upsilon_{1}\ \Longrightarrow\ \mu^{*} = \frac{1}{1+\Upsilon_{1}}\ ,\ \  \varepsilon = 1-\frac{\mu^{*}}{\mu}\ . \label{crossoverpoint}
\end{equation}
By definition, $A_{\mu^{*}} = 0$ and $\eta_{\mu^{*}} = \frac{\pi}{4}$. We also have
\begin{eqnarray}
\cot 2\eta_{\mu} &=& \frac{1}{2\alpha \sqrt{\Upsilon_{2}}}\left(\frac{1-\mu^{*}}{\mu^{*}}-\frac{1-\varepsilon}{\mu^{*}}+1\right) \nonumber \\
&=& \frac{(1+\Upsilon_{1})}{2\alpha \sqrt{\Upsilon_{2}}}\varepsilon.  \label{cot2etamuexpress}
\end{eqnarray}
The bound $\Upsilon_{2} \leq \xi_{1}^{-2}$ and the assumption $\alpha \ll \xi_{1}$ give $\alpha \sqrt{\Upsilon_{2}} \leq \alpha/\xi_{1} \ll 1$. Then $|\cot 2\eta_{\mu}|$ is large for $\varepsilon$ not close to $0$. On the other hand, for $\mu$ close to $\mu^{*}$,
\begin{displaymath}
|\varepsilon| \ll 1:\ \cot 2\eta_{\mu} = \omega(\mu-\mu^{*})\ ,
\end{displaymath}
\begin{equation}
\omega = \frac{(1+\Upsilon_{1})^{2}}{2\alpha\sqrt{\Upsilon_{2}}}\ \geq\ \frac{\xi_{1}}{2\alpha}\ \gg\ 1\ . \label{cot2etamusmall}
\end{equation} 

From (\ref{adiabaticsolutions}), we obtain the excitation gap as
\begin{eqnarray}
g_{\mu} &=& E_{+,\mu} - E_{-,\mu} = \frac{\alpha(1-\mu)}{\sqrt{\Upsilon_{2}}}(\tan \eta_{\mu}+\cot\eta_{\mu}) \nonumber \\
&=& \frac{2\alpha(1-\mu)}{\sqrt{\Upsilon_{2}}} \csc2\eta_{\mu}\ .   \label{energygap}
\end{eqnarray}
Since $\csc 2\eta_{\mu} \geq |\cot2\eta_{\mu}| \gg 1$ for $\mu$ not close to $\mu^{*}$, $g_{\mu}$ is close to its minimum only when $\mu$ is sufficiently close to $\mu^{*}$. The size of this region is characterized by the parameter $\omega$. Explicitly, using (\ref{cot2etamusmall}) for $\mu - \mu^{*} \ll 1$, we get
\begin{eqnarray}
g_{\mu} &=& \frac{2\alpha(1-\mu^{*}-(\mu - \mu^{*}))}{\sqrt{\Upsilon_{2}}}\sqrt{1+\omega^{2}(\mu-\mu^{*})^{2}} \nonumber \\
&\approx& \frac{2\alpha(1-\mu^{*})}{\sqrt{\Upsilon_{2}}}\left[1-\frac{\mu-\mu^{*}}{1-\mu^{*}}+\frac{\omega^{2}}{2}(\mu-\mu^{*})^{2}\right]. \nonumber
\end{eqnarray}
At its minimum,
\begin{displaymath}
\mu_{\rm min} = \mu^{*}+\frac{1}{\omega^{2}(1-\mu^{*})}\ ,
\end{displaymath}
\begin{equation}
g_{\rm min} =  \frac{2\alpha(1-\mu^{*})}{\sqrt{\Upsilon_{2}}}\left[1-\frac{1}{2\omega^{2}(1-\mu^{*})^{2}}\right]\ . \label{minimumgap}
\end{equation}
With $\omega \gg 1$, the deviations of these values from their values at the crossover point are tiny. The assumption $\xi_{N-1} \not\gg 1$ gives $\sqrt{\Upsilon_{2}}\ \geq\ \xi_{N-1}^{-1} = \Omega(1)$, and hence $g_{\rm min} = O(\alpha)$. Also, for $\mu$ close to $\mu^{*}$, (\ref{energygap}) and (\ref{minimumgap}) can be combined as $g_{\mu} = g_{\rm min}\csc 2\eta_{\mu}$.  

As $g_{\rm min} = O(\alpha)$, (\ref{adiabatictime}) implies that the time required for the standard adiabatic evolution to be successful is $\Gamma \geq O(\alpha^{-2})$. We observe that the state $|t\rangle$ can also be obtained by a simple scheme of $O(\alpha^{-2})$ times preparation and subsequent measurements of the state $|s\rangle$ in a suitable basis. Hence, the standard adiabatic evolution does not give any speedup over the simple scheme. In the next section, we show that if we know the crossover point $\mu^{*}$ then we can achieve a faster algorithm with the time complexity $O(\alpha^{-1})$.
  
\section{Partial adiabatic evolution}

Before presenting the faster algorithm, we first compute the overlap of the ground state $E_{-,\mu}$ of $\mathsf{H}_{\mu}$ with the initial and final ground states, $|s\rangle$ and $|t\rangle$. With the normalization condition $\sum_{\ell}|\langle \ell|E_{-,\mu}\rangle|^{2} = 1$, (\ref{lekuoverlap}) gives
\begin{equation}
\sum_{\ell} \frac{|\langle \ell|t\rangle|^{2}}{[\xi_{\ell}(1-\mu)-E_{-,\mu}]^{2}} = \frac{1}{\mu^{2}|\langle t|E_{-,\mu}\rangle|^{2}}\ .
\end{equation} 
For $E_{-,\mu} \ll \xi_{\ell}(1-\mu)$, $O(E_{-,\mu}/\xi_{\ell}(1-\mu))$ terms can be ignored in above equation to get
\begin{equation}
\frac{\alpha^{2}}{E_{-,\mu}^{2}}+\sum_{\ell \neq 0}\frac{|\langle \ell|t\rangle|^{2}}{\xi_{\ell}^{2}(1-\mu)^{2}} = \frac{1}{\mu^{2}|\langle t|E_{-,\mu}\rangle|^{2}} 
\end{equation}
or
\begin{equation}
\frac{\alpha^{2}}{E_{-,\mu}^{2}} + B_{\mu}^{2} = \frac{1}{\mu^{2}|\langle t|E_{-,\mu}\rangle|^{2}}\ .
\end{equation}
where we have used the definition of $B_{\mu}$ (\ref{adiabaticcoeff}). Using (\ref{adiabaticsolutions}) for $E_{-,\mu}$, above equation gives
\begin{displaymath}
\frac{B_{\mu}^{2}}{\cos^{2}\eta_{\mu}} = \frac{1}{\mu^{2}|\langle t|E_{-,\mu}\rangle|^{2}}\ \Longrightarrow\ |\langle t|E_{-,\mu}\rangle| = \frac{1}{\mu}\frac{\cos \eta_{\mu}}{B_{\mu}} 
\end{displaymath} 
or
\begin{equation}
|\langle t|E_{-,\mu}\rangle| = [({\textstyle \frac{1}{\mu}}-1)/\sqrt{\Upsilon_{2}}] \cos \eta_{\mu}\ . \label{tground}
\end{equation}
To compute $|\langle s|E_{-,\mu}\rangle|$, we put $\ell = 0 = s$, $\xi_{0} = 0$, (\ref{adiabaticsolutions}) and (\ref{tground}) in (\ref{lekuoverlap}) to get
\begin{equation}
|\langle s|E_{-,\mu}\rangle| = \mu\frac{\alpha}{|E_{-,\mu}|}\frac{|\cos \eta_{\mu}|}{B_{\mu}}  = \sin \eta_{\mu}\ .  \label{sground}
\end{equation}
 
Now, consider the narrow interval $[\mu^{-},\mu^{+}]$, where $\mu^{\pm} = \mu^{*} \pm c\omega^{-1}$ with $1 \ll c \ll \omega$. Outside this interval, we have $|\mu - \mu^{*}| \geq c\omega^{-1}$, and $|\cot 2\eta_{\mu}| \geq c$ from (\ref{cot2etamusmall}). Therefore, $\eta_{\mu \geq \mu^{+}}$ is close to zero, $\eta_{\mu \leq \mu^{-}}$ is close to $\frac{\pi}{2}$, and $g_{\mu} = g_{\rm min}\csc 2\eta_{\mu}$ is much larger than $g_{\rm min}$ outside $[\mu^{-},\mu^{+}]$. We will also find below that the ground state $|E_{-,\mu}\rangle$ of $\mathsf{H}_{\mu}$ changes substantially only within the interval $[\mu^{-},\mu^{+}]$. This property can be used to construct a faster adiabatic algorithm which performs a partial adibatic evolution only within the interval $[\mu^{-},\mu^{+}]$ and safely skips the evolution outside this interval.  

The validity of our analysis relies on (\ref{assumption}). With $g_{\mu} = |E_{+,\mu}| + |E_{-,\mu}|$ and (\ref{energygap}), this validity condition holds provided $\alpha \csc 2\eta_{\mu} \ll \xi_{1}\sqrt{\Upsilon_{2}}/2$. Now for $\mu \in [\mu^{-},\mu^{+}]$, $\alpha \csc 2\eta_{\mu} \leq \alpha \sqrt{1+c^2}$, and $\xi_{1}\sqrt{\Upsilon_{2}}/2 \geq \xi_{1}/2\xi_{N-1} \not\ll 1$ due to the assumption (\ref{firstassumption}). Since $\alpha \ll \xi_{1}$, the validity condition can be satisfied by keeping $c$ small compared to $\xi_{1}/\alpha$. That also keeps the interval $[\mu^{-},\mu^{+}]$ narrow, with $c\ll\omega$ as per (\ref{cot2etamusmall}). Thus our analysis is valid within the interval $[\mu^{-},\mu^{+}]$.

Evolution of $|E_{-,\mu}\rangle$ is obtained by (\ref{tground}) and (\ref{sground}), i.e.
\begin{eqnarray}
&|\langle s|E_{-,\mu}\rangle| = \sin \eta_{\mu}\ ,& \nonumber \\
&|\langle t|E_{-,\mu}\rangle| = [({\textstyle \frac{1}{\mu}}-1)/\sqrt{\Upsilon_{2}}] \cos \eta_{\mu}\ .& \label{adiabaticinitialprojection}
\end{eqnarray}
We have $\eta_{\mu}\approx\frac{\pi}{2}$ for $\mu \leq \mu^{-}$, and $\eta_{\mu} \approx 0$ for $\mu \geq \mu^{+}$. So in passing through the interval $[\mu^{-},\mu^{+}]$, the ground state $|E_{-,\mu}\rangle$ transforms from being very close to the initial state $|s\rangle$ to being almost orthogonal to $|s\rangle$. (Note that when $\eta_{\mu}$ is a smooth function, $|E_{-,\mu}\rangle$ is close to $|s\rangle$ for all $\mu\leq\mu^{-}$, even though our analysis does not hold for all $\mu\leq\mu^{-}$.) Simultaneously, the overlap of $|E_{-,\mu}\rangle$ with the target state $|t\rangle$ increases from zero to
\begin{eqnarray}
\langle t|E_{-,\mu^{+}}\rangle&\approx&({\textstyle \frac{1}{\mu^{+}}}-1)/\sqrt{\Upsilon_{2}}\ \nonumber \\
 &\approx&({\textstyle \frac{1}{\mu^{*}}}-1)/\sqrt{\Upsilon_{2}}=\Upsilon_{1}/\sqrt{\Upsilon_{2}}\ .  \label{tEmu+}
\end{eqnarray}  
We have $1 \geq \Upsilon_{1}/\sqrt{\Upsilon_{2}} \geq \xi_{1}/\xi_{N-1}$ due to the bounds mentioned after (\ref{adiabaticOmega}). The assumption $\xi_{1}/\xi_{N-1} \not\ll 1$ then implies that $|t\rangle$ has a significant overlap with $|E_{-,\mu^{+}}\rangle$. Hence $|t\rangle$ can be obtained by few preparations and subsequent measurements of $|E_{-,\mu^{+}}\rangle$.

We now define the partial adiabatic evolution as evolution from $\mathsf{H}_{\mu^{-}}$ to $\mathsf{H}_{\mu^{+}}$, as opposed to the complete adiabatic evolution from $\mathsf{H}_{s}$ to $\mathsf{H}_{t}$. The resultant algorithm executes the three steps below:\\
(1) The initial state $|s\rangle$ is prepared as the stable ground state of the Hamiltonian $\mathsf{H}_{s}$. At $\tau=0$, the Hamiltonian is suddenly changed to $\mathsf{H}_{\mu^{-}}$, without disturbing the state $|s\rangle$~\cite{messiah}. The system is then in the ground state $|E_{-,\mu^{-}}\rangle$ with probability $\sin^{2}\eta_{\mu^{-}}$.\\
(2) The Hamiltonian evolves from $\mathsf{H}_{\mu^{-}}$ to $\mathsf{H}_{\mu^{+}}$, linearly in time over duration $\Gamma$. The system encounters the minimum excitation gap $g_{\rm min}$ during this evolution, and the state $|E_{-,\mu^{-}}\rangle$ reaches the state $|E_{-,\mu^{+}}\rangle$ with probability close to $1$ for $\Gamma\geq 2c\omega^{-1}g_{\rm min}^{-2}$.\\
(3) The state of the system is measured. The state $|E_{-,\mu^{+}}\rangle$ yields the target state $|t\rangle$ with probability $\Upsilon_{1}^2/\Upsilon_{2}$. These three steps are repeated until we find $|t\rangle$.

The combined success probability of the three steps is $P_{\rm ad} = \sin^{2}\eta_{\mu^{-}} \Upsilon_{1}^{2}/\Upsilon_{2}$.
The overall complexity of the algorithm is, to leading order,
\begin{equation}
\Gamma' = \frac{\Gamma}{P_{\rm ad}} \geq \frac{2c}{\omega g_{\rm min}^{2}} \times \frac{\csc^{2}\eta_{\mu^{-}}\Upsilon_{2}}{\Upsilon_{1}^{2}} = \frac{c}{\alpha}\frac{\Upsilon_{2}^{5/2}}{\Upsilon_{1}^{4}} \left(1+\frac{1}{4c^{2}}\right) \ .  \label{fasterevolutiontime} 
\end{equation}
making use of (\ref{cot2etamusmall}) and (\ref{minimumgap}). Since $\Upsilon_{p} \in [\xi_{N-1}^{-p},\xi_{1}^{-p}]$, and we have assumed both $\xi_{1}$ and $\xi_{N-1}$ to be $\Theta(1)$, the factor $\Upsilon_{2}^{5/2}/\Upsilon_{1}^{4}$ is also $\Theta(1)$. That makes $\Gamma' \geq \Theta(\alpha^{-1})$, and hence the partial adiabatic evolution is indeed quadratically faster than the $\Theta(\alpha^{-2})$ classical search algorithms.

\section{Discussion}

We can obtain Roland and Cerf's results~\cite{rolandlocal} as a special case of our partial adiabatic algorithm. There $\mathsf{H}_{s} = \mathbbm{1}_{N}-|u\rangle\langle u|$, $|u\rangle = \sum_{j}|j\rangle/\sqrt{N}$ and $\alpha = 1/\sqrt{N}$. We then have $\xi_{\ell \neq s} = 1$ and $\Upsilon_{1} = \Upsilon_{2} = \sum_{\ell \neq s}|\langle \ell|t\rangle|^{2} = 1-\alpha^{2}$. It follows that the crossover point is $\mu^{*} \approx 1/2$, with $\omega \approx 2/\alpha$ and $g_{\rm min}\approx\alpha$. The width of the narrow interval $[\mu^{-},\mu^{+}]$ is $2c\omega^{-1} = c\alpha \ll 1$ as desired. Using the partial adiabatic evolution algorithm, we obtain the target state in time $\Gamma' \geq c\Upsilon_{2}^{5/2}/\alpha \Upsilon_{1}^{4} \approx c\sqrt{N}$, which is optimal up to a constant factor. Roland and Cerf obtained the optimal algorithm by performing local adiabatic evolution which performs the evolution slowly around the crossover point where the energy gap is small. 
Note that the knowledge of the crossover point $\mu^{*}$ is essential to get the optimal algorithm, either by partial evolution or local evolution. Another special case of our analysis is due to Farhi et al.~\cite{farhiqcae}, where $\mathsf{H}_{s}$ is a sum of single qubit Hamiltonians. 

By time reversal symmetry, our analysis can be extended to the problem where $\mathsf{H}_{s} = -|s\rangle\langle s|$ and $\mathsf{H}_{t}$ is a general Hamiltonian. The required interchanges are $t \leftrightarrow s$ and $\mu \rightarrow 1-\mu$. Also, $\Upsilon_{p} = \sum_{j\neq t}|\langle j|s\rangle|^{2}/\xi_{j}^{p}$, where $\{|j\rangle,\xi_{j}\}$ represent the eigenspectrum of $\mathsf{H}_{t}$ with $\xi_{j=t} = 0$. The crossover point becomes $\mu^{*} = \Upsilon_{1}/(1+\Upsilon_{1})$, and $\{g_{\mu},\omega,\Gamma\}$ can be calculated. \v{Z}nidari\v{c} and Horvat~\cite{znidaric} have studied a particular case of this type, with $\mathsf{H}_{t}$ representing instances of an NP-complete problem.

\textbf{Acknowledgments}: I thank Prof. Apoorva Patel for going through the manuscript and for useful comments and discussions.

\end{document}